# Can tumor location on pre-treatment MRI predict likelihood of pseudo-progression versus tumor recurrence in Glioblastoma? – A feasibility study

**Marwa Ismail[1], Virginia Hill[2], Volodymyr Statsevych[2], Evan Mason[2], Ramon Correa[1], Prateek Prasanna[1], Gagandeep Singh[1], Kaustav Bera[1], Rajat Thawani[1], Anant Madabhushi[1], Manmeet Aahluwalia[3], and Pallavi Tiwari[1].**

[1]Department of Biomedical Engineering, Case Western Reserve University, Cleveland, OH, USA

[2]Department of Neuroradiology, Imaging Institute, Cleveland Clinic, Cleveland, OH, USA

[3]Brain Tumor and Neuro-Oncology Center, Cleveland Clinic, Cleveland, OH, USA

*** Correspondence:**
Marwa Ismail
mxi125@case.edu

**Abstract**

A significant challenge in Glioblastoma (GBM) management is identifying pseudo-progression (PsP), a benign radiation-induced effect, from tumor recurrence, on routine imaging following conventional treatment. Previous studies have linked tumor lobar presence and laterality to GBM outcomes, suggesting that disease etiology and progression in GBM may be impacted by tumor location. Hence, in this feasibility study, we seek to investigate the following question: Can tumor location on treatment-naïve MRI provide early cues regarding likelihood of a patient developing pseudo-progression versus tumor recurrence? In this study, 74 pre-treatment Glioblastoma MRI scans with PsP (33) and tumor recurrence (41) were analyzed. First, enhancing lesion on Gd-$T_{1w}$ MRI and peri-lesional hyperintensities on $T_{2w}$/FLAIR were segmented by experts and then registered to a brain atlas. Using patients from the two phenotypes, we construct two atlases by quantifying frequency of occurrence of enhancing lesion and peri-lesion hyperintensities, by averaging voxel intensities across the population. Analysis of differential involvement was then performed to compute voxel-wise significant differences (p-value<0.05) across the atlases. Statistically significant clusters were finally mapped to a structural atlas to provide anatomic localization of their location. Our results demonstrate that patients with tumor recurrence showed prominence of their initial tumor in the parietal lobe, while patients with PsP showed a multi-focal distribution of the initial tumor in the frontal and temporal lobes, insula, and putamen. These preliminary results suggest that lateralization of pre-treatment lesions towards certain anatomical areas of the brain may allow to provide early cues regarding assessing likelihood of occurrence of pseudo-progression from tumor recurrence on MRI scans.

## 1 Introduction

A significant challenge in management of Glioblastoma (GBM), the most aggressive form of brain cancer, is differentiating tumor recurrences from pseudo-progression (PsP) on routine magnetic resonance (MR) scans [1]. PsP is a benign radiation-induced treatment effect which occurs in approximately 19 – 33% of all malignant brain tumors [2] and usually stabilizes or regresses without further treatment. Unfortunately, PsP mimics tumor recurrence radiologically on routine MRI scans (Gadolinium-enhanced T1-weighted (Gd-T1w), T2-weighted (T2w), FLAIR), making it challenging to differentiate from true tumor recurrence [2]. Studies have previously explored advanced imaging modalities such as perfusion imaging [3 - 5], MR spectroscopy [4], and diffusion-weighted imaging [5] in distinguishing tumor recurrence from PsP. However, these advanced imaging modalities are limited by acquisition variability, costs, reproducibility, and unavailability at most clinical sites [6]. Reliable disease assessment using routine imaging is thus needed in order to aid in accurately identifying PsP from tumor recurrence. Timely identification of these conditions could avoid unnecessary interventions in patients with PsP, while allowing for change in treatment for patients with tumor recurrence [1].

Multiple studies have linked initial lesion location in the brain to be a prognostic marker of tumor recurrence and overall survival in diffuse Gliomas [7]. For instance, recent studies have demonstrated a higher rate of 1p19q deletion in the frontal lobe [8], and absence of 1DH1 mutation within the insula [9]. Similarly, Gliomas in the frontal locations have been shown to be associated with a better prognosis compared to other locations [10]. Further, enhancing lesion developing in the periventricular region has been linked to PsP [11, 12]. These studies seem to suggest that the underlying disease etiology may be driven by tumor location. Hence, it may be reasonable to rationalize that initial GBM location in the brain may implicitly contribute to an increased likelihood of a patient developing pseudo-progression or tumor recurrence, following conventional treatment of maximal surgical resection and chemo-radiation therapy.

In this feasibility study, we evaluate this hypothesis that lesion location on pre-treatment MR scans could provide early cues regarding likelihood of a patient developing tumor recurrences versus PsP. In order to anatomically localize the disease, we employ "population atlases" of GBM phenotypes to establish predisposition of tumor recurrence or PsP to specific spatial locations in the brain based on their frequency of occurrence [7, 13, 14]. The statistical population atlases allow for the succinct encapsulation of structural and anatomical variability of the disease across a patient population using a single reference or canonical representation. We will construct population atlases on a cohort of 74 brain MRI scans across two lesion sub-compartments (peritumoral hyperintensities as defined on FLAIR scans and enhancing core as defined on $T_{1w}$ MRI), to quantify the frequency of occurrence of PsP and tumor recurrence in pre-treatment lesions. We will further employ a statistical mapping technique, ADIFFI, to identify if there exist any statistically significant lesion locations in the brain across the two disease pathologies, by comparing the population atlases of PsP and tumor recurrence.

## 2 Materials and Methods

### 2.1 Study Population

The Institutional Review Board-approved and HIPAA-compliant study comprised GBM patient population from Cleveland Clinic. The population cohort for pre-treatment cases included 74 cases in total; 41 tumor recurrence cases, and 33 PsP cases. The studies were identified by performing a retrospective review of all brain tumor patients who received chemo-radiation treatment using the Stupp protocol at the respective institutions and had a suspected enhancing lesion within 3 months of treatment. All cases were confirmed for disease presence using the criteria provided below. Informed consent was obtained for all patients involved in the study. All MR scans were acquired using either a 1.5 Tesla or a 3-Tesla scanner. Table 1 summarizes the demographics for this study population.

### 2.2 Confirmation for disease presence

Our inclusion criteria consisted of the following: (1) pre-, and post-treatment MRI scans that are of diagnostic image quality as determined by collaborating radiologists, (2) availability of all 3 routine MRI sequences (Gd-T1w, T2w, FLAIR), (3) a suspected post-treatment enhancing lesion with more than 5 millimeters (mm) of rim or nodular enhancement, and (4) confirmation of PsP or tumor recurrence for the suspected lesion, obtained using RANO criteria [15] on follow-up MRI scans.

### 2.3 Image Registration and Tumor segmentation

Every pre-treatment MRI lesion was annotated into 2 regions: enhancing lesion and $T_{2w}$/FLAIR hyperintense peri-lesional component. Gd-$T_{1w}$ MRI scans were used to delineate the enhancing lesion, while both $T_{2w}$ and FLAIR scans were used to annotate the $T_{2w}$/FLAIR hyperintense peri-lesional compartment. All annotations were performed by two experienced readers via an open source hand-annotation tool in 3D Slicer [16].

In order to map all scans to the same space for the purpose of spatial atlas construction, the 3 MRI sequences for each patient, Gd-$T_{1w}$ MRI, $T_{2w}$, and FLAIR, were co-registered to a 1.0-mm isotropic T1-weighted brain atlas (MNI152; Montreal Neurological Institute) using mutual information with 12-degrees of freedom. This was followed by visual inspection to make sure all images were properly aligned. Skull stripping was then performed using a deformable surface classification algorithm [17], followed by bias field correction that was performed using the nonparametric non-uniform intensity normalization technique in [18].

### 2.4 Frequency Map Construction

From the available annotations for both enhancing lesion and $T_{2w}$/FLAIR hyperintense peri-lesional compartments, population atlases for each compartment were built for both pathologies (tumor recurrence and PsP). These atlases were constructed to quantify the frequency of occurrence of both enhancing lesion and peri-lesional hyperintensities across tumor recurrence and PsP, by averaging intensity values for all voxels across all the annotated binary images of all patients involved in the study. The frequency of lesion occurrence was visualized using a heat map superimposed on the reference MNI152 atlas.

### 2.5 Analysis of Differential Involvement (ADDIFI)

From the constructed tumor progression and PsP frequency atlases, analysis of differential involvement (ADIFFI) was performed as described in [7], once for the enhancing lesion compartment and once for the peri-lesional hyperintensities. First, a two-tailed Fisher's exact test was conducted, to evaluate a 2x2 contingency table that compares tumor recurrence/PsP along with tumor/non-tumor occurrence for each voxel across all patients. From this voxel-wise analysis, significance level was then measured, and the voxels that yielded p-value<0.05 were stored. The voxel-wise probabilities according to Fisher's exact test are computed using the following formula:

$$\boldsymbol{p} = \frac{(\boldsymbol{a}+\boldsymbol{b})!\,(\boldsymbol{c}+\boldsymbol{d})!\,(\boldsymbol{a}+\boldsymbol{c})!\,(\boldsymbol{b}+\boldsymbol{d})!}{\boldsymbol{a}!\,\boldsymbol{b}!\,\boldsymbol{c}!\,\boldsymbol{d}!\,\boldsymbol{n}!},$$

where $a, b, c, d,$ and $n$ are defined as follows:

$a$: represents the number of tumor recurrence as well as the lesion-positive occurrences across all subjects at the current voxel.

$b$: represents the number of tumor progression as well as the lesion-negative occurrences across all subjects at the current voxel.

$c$: represents the number of PsP as well as the lesion-positive occurrences across all subjects at the current voxel.

$d$: represents the number of PsP as well as the lesion-negative occurrences across all subjects at the current voxel.

$n$: represents the total number of studies.

Next, connected component analysis was applied, [19], to cluster all statistically significant voxels found across the two compartments for both tumor recurrence and PsP that appeared on the ADIFFI maps, for enhancing lesion as well as for peri-lesional hyperintensities. The brain was finally partitioned using pre-labeled anatomical structures in MNI space [20], for the purpose of identifying the anatomic areas of localization for tumor recurrence/PsP across all subjects.

### 2.6 Cluster-size correction using random permutation analysis

Due to the extensive number of voxel-wise calculations performed during ADIFFI, multiple comparison corrections were performed. For this task, random permutation (RP) analysis was conducted for cluster size correction [21]. Specifically, all $T_{2w}$/FLAIR hyperintense peri-lesional components, as well as the enhancing lesion ones, across the two categories (tumor recurrence/ PsP) were randomly reassigned to one of these pathologies, then ADIFFI was re-conducted, and voxels with p-values less than 0.05 were stored. In addition, the sizes of statistically significant clusters were documented. The whole process was reiterated for 500 times. RP analysis was employed in order to identify distinct clusters occurring less than 5% by chance, which would provide distinct spatial differences between tumor recurrence and PsP.

Finally, statistically significant clusters appearing on the cluster-size corrected ADIFFI maps were designated as either PsP or tumor recurrence by referring to the population atlases that were individually constructed for tumor recurrence and PsP. A specific anatomic localization was then obtained from these cluster-size corrected ADIFFI maps, by mapping them to a structural MNI atlas. The entire pipeline of this work is shown in Figure [1].

## 3 Results

The resulting frequency maps that were constructed for both T2w/FLAIR hyperintense peri-lesional and lesion areas from pre-treatment scans are shown in Figures [2], [3] respectively. These figures show that tumor recurrence in both compartments (enhancing lesion and T2w/FLAIR hyperintense peri-lesional areas) is more likely lateralized towards the parietal lobe, whereas PsP is more likely to be multi-focally distributed across different anatomical areas of the brain including frontal and temporal lobes, the insula, and the putamen.

### 3.1 Tumor recurrence is lateralized towards the parietal lobe

The frequency maps as well as ADIFFI maps for peri-lesional T2/FLAIR hyperintensities of the pre-treatment scans show that tumor recurrence is more likely to be present in the parietal lobe, with frequency of occurrence of 85%, Fig [2 a], Fig [4 a]. About 59% of this distribution was found in the right hemisphere, whereas 41% was found in the left hemisphere. Frequency maps as well as ADIFFI maps obtained for the enhancing lesion also reveal that tumor recurrence is more likely to be present in the parietal lobe of left and right hemispheres (70% and 30% chances of occurrence respectively), Fig [3 a], Fig [4 c]. These results suggest that tumor recurrence exhibits lobar prominence across the population atlases, but do not exhibit any hemisphere-specific preference.

### 3.2 Pseudo-progression exhibits a multi-focal distribution in the enhancing lesion as well as the perilesional hyperintensities

PsP, unlike tumor recurrence, seems to more likely be multi-focally distributed across the brain in pre-treatment cases, for both the enhancing lesion and the peri-lesional hyperintensities. PsP exhibited a multi-focal distribution in the right hemisphere of the peri-lesional hyperintensities, with frequencies of occurrence of 55% in the frontal lobe, 11% in the temporal lobe, 10% in the insula, and 10% in the putamen (Fig [2 b], Fig [4 b]). In the analysis of the enhancing lesion regions, PsP appears to more likely be multi-focally distributed within both left and right hemispheres. The spatial distribution was 35% in the insula (with 63% of this distribution in the right hemisphere and 37% in the left hemisphere), 21% in the right frontal lobe, 13% in the right temporal lobe, and 17% in the putamen (with 57% of this distribution in the right hemisphere and 43% in the left hemisphere), Fig [3 b], Fig [4 d].

### 3.3 Random permutation analysis for cluster size correction

RP analysis conducted on the peri-lesional T2/FLAIR hyperintensities of the pre-treatment cases revealed that the average and standard deviation of maximum cluster size are 3700 and 1726.8 voxels respectively. Also, 95% of the cluster sizes were smaller than 6192 voxels, meaning that clusters larger than this size threshold would occur in less than 5% of all random permutations. This resulted in one distinct T2w/FLAIR hyperintense peri-lesional cluster size of 6502 voxels, localized at the right parietal lobe, and associated with tumor recurrence, and another one of size of 6200 voxels localized at the left parietal lobe.

RP analysis conducted on the enhancing lesion revealed that average and standard deviation of maximum cluster size are 2258 and 1774.1 voxels respectively. Also, 95% of the cluster sizes were smaller than 5164 voxels, meaning that clusters larger than this size threshold would occur in less than 5% of all random permutations. This resulted in one distinct enhancing lesion cluster size of 5450 voxels, localized at the left parietal lobe, and associated with tumor recurrence.

The designation of PsP or true progression based on ADIFFI maps as for each significant voxel/cluster was accomplished by referring to the population atlases of both compartments (enhancing lesion, T2w/FLAIR hyperintense peri-lesion) that were individually constructed for tumor recurrence and PsP. The cluster-size corrected ADIFFI maps obtained for tumor recurrence are shown in Fig [1 d].

The extent of resection (available for n=37 subjects), age, and gender were not found to be independently prognostic of presence of PsP versus tumor recurrence.

## 4 Discussion

Distinguishing tumor recurrence from PsP is one of the biggest clinical challenges in GBM management. This feasibility study aimed at creating population atlases to study spatial proclivity of brain tumor recurrence versus PsP based on their occurrences on pre-treatment MR scans. The study assessed the voxel-wise tumor frequency across two lesion compartments using a statistical mapping technique named ADIFFI, in efforts to find significant spatial distribution differences between the two phenotypes.

Our preliminary findings suggest that likelihood of tumor recurrence is more consistent with lesions occurring in the parietal lobe of both left and right hemispheres, based on the analysis of both enhancing lesion and peri-lesional T2/FLAIR hyperintensities, on pre-treatment MRI scans. Parietal lobe is largely responsible for cognitive functions. Damage to parietal lobe may have direct implications in processing speech as well as sensory information. Hence, presence of tumor recurrence in parietal lobe may cause symptoms associated with numbness and tingling, hemi-neglect, and cognitive issues around right-left confusion and reading and math problems. PsP, on the other hand, did not exhibit lobar-specific distribution in pre-treatment scans, but showed a multi-focal distribution of the initial tumor in the frontal (associated with motor function, memory, problem solving) and temporal lobes (associated with primary auditory perception, such as hearing and visual recognition) as well as the insula and putamen. While the association of presence of tumor recurrence or PsP with specific lobes in the brain is not well-understood, their presence in specific lobes could ultimately contribute towards making more informed decisions regarding their diagnosis.

Previous studies have largely employed population atlases in brain tumors using pre-treatment MRI to obtain probabilistic maps of spatial predisposition in patients based on their disease aggressiveness [22] or molecular status [7, 23, 24]. For instance, a few studies have shown that tumor recurrence closer to the ventricular system was significantly associated with poor survival [25, 26]. Interestingly, the study in [27] showed that tumors in the right occipito-temporal periventricular white matter were significantly associated with poor survival in both training and test cohorts. Similarly, more aggressive GBMs were reported to be close to the ventricular system, and had a rapid progression [28], suggesting that tumor location may play a significant role in disease etiology.

The closest studies to our work have attempted to identify associations of lesion location with likelihood of tumor recurrence and PsP, to investigate any spatial differences between the two phenotypes. For instance, the study by Tsien et al. [29] incorporated location along with clinical and conventional MRI parameters to distinguish tumor progression from PsP in high-grade gliomas, yet no significant location differences could be found between the two groups, perhaps on account of the relatively small population size involved in this study (27 patients total). The study by Van West et al [11] reported the incidence of PsP in low grade gliomas, and found that 50% of their PsP enhancing lesions were located in the periventricular walls; attributing to the relatively poor blood supply in the periventricular areas that make it more vulnerable to radiation-induced processes. However, these studies did not report any findings regarding lobular preferences for either PsP or tumor recurrence in GBMs.

Our study did have its limitations. While our results are promising as a feasibility study, the study did not consider the molecular status (i.e. MGMT), or Karnofsky performance score as potential confounders during analysis. While the extent of resection on a limited subset of studies was not found to be independently prognostic of tumor recurrence, these findings need to be validated on a larger multi-institutional cohort. The prognostic implications (i.e. predicting patient overall survival), based on the location differences across PsP and tumor recurrence were not studied as a part of the current work, and will be investigated in the future.

To conclude, this study attempted to demonstrate the likelihood of occurrence of tumor recurrence and pseudo-progression, using the location of the lesion on pre-treatment MR scans. Our results revealed distinct localization between tumor recurrences and PsP that could aid in predicting these two similar appearing pathological conditions. Future work will focus on integrating the location biomarker with other biomarkers, such as shape and texture features, on a larger cohort of multi-institutional studies. We will also consider identifying location specific markers associated with radiation necrosis (delayed treatment effects) versus tumor recurrence.

## 5 Conflict of Interest

*The authors declare that the research was conducted in the absence of any commercial or financial relationships that could be construed as a potential conflict of interest*.

## 6 Author Contributions

M.I and P.T contributed to analysis and interpretation of data. M.I and P.T contributed to designing the experiments, drafting, and revising the article. V.H, V.S, E.M provided clinical datasets, and interpretation of radiographic images. M.A helped define the clinical problem and provided clinical interpretation of findings. V.H, R.C, P.P, G.S, K.B, R.T curated the studies and performed the annotations on radiological images. A.M revised the manuscript critically for important intellectual content. All authors have reviewed the manuscript.

## 7 Funding

Research reported in this publication was supported by the National Cancer Institute of the National Institutes of Health under award numbers 1U24CA199374-01, R01CA202752-01A1, R01CA208236-01A1, R01 CA216579-01A1, R01 CA220581-01A1, National Center for Research Resources under award number 1 C06 RR12463-01, the DOD Prostate Cancer Idea Development Award; the DOD Lung Cancer Idea Development Award; Dana Foundation David Mahoney Neuroimaging Program, the Ohio Third Frontier Technology Validation Fund, the Wallace H. Coulter Foundation Program in the Department of Biomedical Engineering and the Clinical and Translational Science Award Program (CTSA) at Case Western Reserve University. The content is solely the responsibility of the authors and does not necessarily represent the official views of the National Institutes of Health.

## 8 Acknowledgments

The authors would like to thank Dr. Benjamin Ellingson for his guidance throughout the statistical experiment.

## 9 References

[1] Parvez K, Parvez A, Zadeh G. The diagnosis and treatment of pseudo-progression, radiation necrosis and brain tumor recurrence. *International journal of molecular sciences* 2014; 15(7):11832-46.

[2] Wang S, Martinez-Lage M, Sakai Y, Chawla S, Kim SG, Alonso-Basanta M, Lustig RA, Brem S, Mohan S, Wolf RL, and Desai A. Differentiating tumor progression from pseudo-progression in patients with glioblastomas using diffusion tensor imaging and dynamic susceptibility contrast MRI. *American Journal of Neuroradiology* 2015; 37 (1): 28 - 36

[3] Detsky JS, Keith J, Conklin J, Symons S, Myrehaug S, Sahgal A, Heyn CC, Soliman H. Differentiating radiation necrosis from tumor progression in brain metastases treated with stereotactic radiotherapy: utility of intravoxel incoherent motion perfusion MRI and correlation with histopathology. *Journal of neuro-oncology 2017*; 134(2):433-41.

[4] Chuang MT, Liu YS, Tsai YS, Chen YC, Wang CK. Differentiating radiation-induced necrosis from recurrent brain tumor using MR perfusion and spectroscopy: a meta-analysis. *PLOS one* 2016; 11(1): e0141438.

[5] Prager AJ, Martinez N, Beal K, Omuro A, Zhang Z, Young RJ. Diffusion and perfusion MRI to differentiate treatment-related changes including pseudo-progression from recurrent tumors in high-grade gliomas with histopathologic evidence. *American Journal of Neuroradiology* 2015; 36(5):877-85.

[6] Brandsma D, Stalpers L, Taal W, Sminia P, van den Bent MJ. Clinical features, mechanisms, and management of pseudo-progression in malignant gliomas. *The Lancet Oncology* 2008; 9: 453-61.

[7] Ellingson BM, Cloughesy TF, Pope WB, Zaw TM, Phillips H, Lalezari S, Nghiemphu PL, Ibrahim H, Naeini KM, Harris RJ, and Lai A. Anatomic localization of O6-methylguanine DNA methyltransferase (MGMT) promoter methylated and unmethylated tumors: a radiographic study in 358 de novo human glioblastomas. *Neuroimage* 2012; 59(2):908-916.

[8] Laigle-Donadey F, Martin-Duverneuil N, Lejeune J, Criniere E, Capelle L, Duffau H, Cornu P, Broët P, Kujas M, Mokhtari K, Carpentier A. Correlations between molecular profile and radiologic pattern in oligodendroglial tumors. Neurology. 2004 Dec 28;63(12):2360-2.

[9] Metellus P, Coulibaly B, Colin C, de Paula AM, Vasiljevic A, Taieb D, Barlier A, Boisselier B, Mokhtari K, Wang XW, Loundou A. Absence of IDH mutation identifies a novel radiologic and molecular subtype of WHO grade II gliomas with dismal prognosis. Acta neuropathologica. 2010 Dec 1;120(6):719-29.

[10] Stockhammer F, Misch M, Helms HJ, Lengler U, Prall F, Von Deimling A, Hartmann C. IDH1/2 mutations in WHO grade II astrocytomas associated with localization and seizure as the initial symptom. Seizure. 2012 Apr 1;21(3):194-7.

[11] Van West SE, de Bruin HG, van de Langerijt B, Swaak-Kragten AT, Van Den Bent MJ, Taal, W. Incidence of pseudo-progression in low-grade gliomas treated with radiotherapy. *Neuro-oncology* 2017; 19(5):719-25.

[12] Patel U, Patel A, Cobb C, Benkers T, Vermeulen S. The management of brain necrosis as a result of SRS treatment for intra-cranial tumors. *Translational Cancer Research* 2014; 3(4):373-82.

[13] Bilello M, Akbari H, Da X, Pisapia JM, Mohan S, Wolf RL, O'Rourke DM, Martinez-Lage M, and Davatzikos C. Population-based MRI atlases of spatial distribution are specific to patient and tumor characteristics in glioblastoma. *NeuroImage: Clinical* 2016; 12: 34-40.

[14] Larjavaara S, Mäntylä R, Salminen T, Haapasalo H, Raitanen J, Jääskeläinen J, Auvinen A. Incidence of gliomas by anatomic location. *Neuro-oncology* 2007; 9(3):319-25.

[15] Wen PY, Macdonald DR, Reardon DA, et al. Updated response assessment criteria for high-grade gliomas: response assessment in neuro-oncology working group. *Journal of clinical oncology* 2010; 28: 1963-72.

[16] Kikinis R, Pieper SD, and Vosburgh KG. 3D Slicer: a platform for subject-specific image analysis, visualization, and clinical support. In *Intraoperative imaging and image-guided therapy* 2014; 277-289. Springer, New York, NY.

[17] Tao X, Chang MC. A skull stripping method using deformable surface and tissue classification. In *International Society for Optics and Photonics Medical Imaging: Image Processing* 2010; 7623: 76233L.

[18] Tustison NJ, Avants BB, Cook PA, Zheng Y, Egan A, Yushkevich PA, and Gee J C. N4ITK: improved N3 bias correction. *IEEE Transactions on Medical Imaging* 2010; 29 (6): 1310-20.

[19] L. Vincent, "Morphological grayscale reconstruction in image analysis: Applications and efficient algorithms," *IEEE transactions on image processing* 2(2), 176–201 (1993).

[20] Mazziotta J, Toga A, Evans A, Fox P, Lancaster J, Zilles K, Woods R, Paus T, Simpson G, Pike B, Holmes C. A probabilistic atlas and reference system for the human brain: International Consortium for Brain Mapping (ICBM). *Philosophical Transactions of the Royal Society of London B: Biological Sciences* 2001; 356(1412):1293-322.

[21] Bullmore ET, Suckling J, Overmeyer S, Rabe-Hesketh S, Taylor E, Brammer MJ. Global, voxel, and cluster tests, by theory and permutation, for a difference between two groups of structural MR images of the brain. *IEEE transactions on medical imaging* 1999; 18(1):32-42.

[22] Duffau H, Capelle L. Preferential brain locations of low-grade gliomas: Comparison with glioblastomas and review of hypothesis. *Cancer* 2004; 100(12):2622-6.

[23] Drabycz S, Roldán G, De Robles P, Adler D, McIntyre JB, Magliocco AM, Cairncross JG, and Mitchell JR. An analysis of image texture, tumor location, and MGMT promoter methylation in glioblastoma using magnetic resonance imaging. *Neuroimage* 2010; 49(2):1398-405.

[24] Kanas VG, Zacharaki EI, Thomas GA, Zinn PO, Megalooikonomou V, Colen RR. Learning MRI-based classification models for MGMT methylation status prediction in glioblastoma. *Computer Methods and Programs in Biomedicine* 2017; 140:249-57.

[25] Adeberg S, König L, Bostel T, Harrabi S, Welzel T, Debus J, Combs SE. Glioblastoma recurrence patterns after radiation therapy with regard to the subventricular zone. *International Journal of Radiation Oncology* Biology* Physics* 2014; 90(4):886-93.

[26] Jafri NF, Clarke JL, Weinberg V, Barani IJ, Cha S. Relationship of glioblastoma multiforme to the subventricular zone is associated with survival. *Neuro-oncology* 2012; 15(1):91-6.

[27] Liu TT, Achrol AS, Mitchell LA, Du WA, Loya JJ, Rodriguez SA, Feroze A, Westbroek EM, Yeom KW, Stuart JM, Chang SD. Computational identification of tumor anatomic location associated with survival in 2 large cohorts of human primary glioblastomas. *American Journal of Neuroradiology* 2016; 37(4): 621-28.

[28] Li HY, Sun CR, He M, Yin LC, Du HG, Zhang JM. Correlation between Tumor Location and Clinical Properties of Glioblastomas in Frontal and Temporal Lobes. *World Neurosurgery* 2018; 112:407-14.

[29] Tsien C, Galbán CJ, Chenevert TL, Johnson TD, Hamstra DA, Sundgren PC, Junck L, Meyer CR, Rehemtulla A, Lawrence T, Ross BD. Parametric response map as an imaging biomarker to distinguish progression from pseudo-progression in high-grade glioma. *Journal of Clinical Oncology* 2010; 28(13):2293.

**Figures**

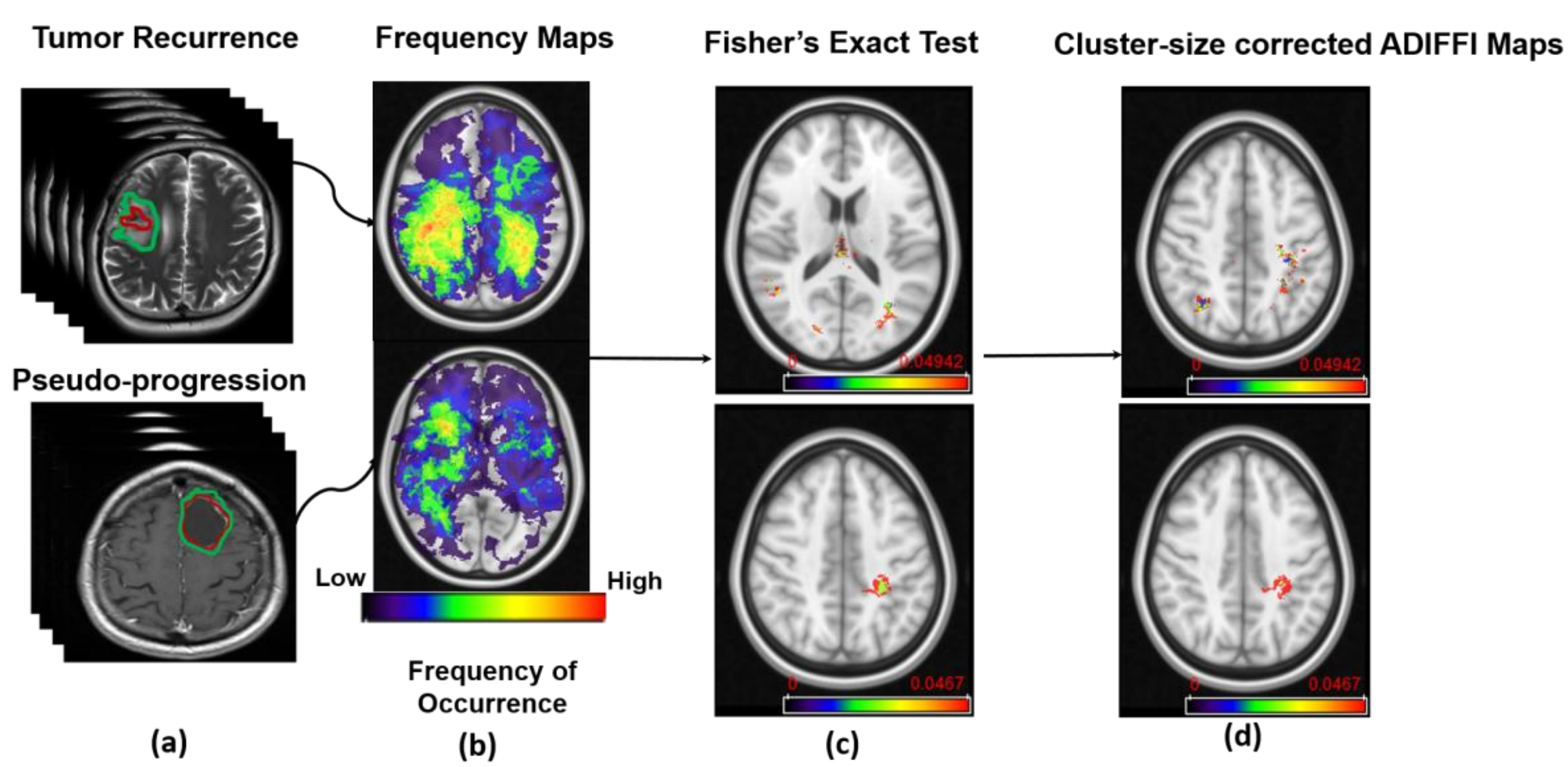

**Figure 1:** Pipeline of the framework. (a) MR scans of tumor recurrence and pseudo-progression. (b) Frequency map atlases that were constructed from the two classes. (c) Results from Fisher's Exact test on peri-lesional T2/FLAIR hyperintensities in tumor recurrence (Top), and enhancing lesion in tumor recurrence (Bottom). (d) Results after applying RP analysis on ADIFFI maps shown in (c).

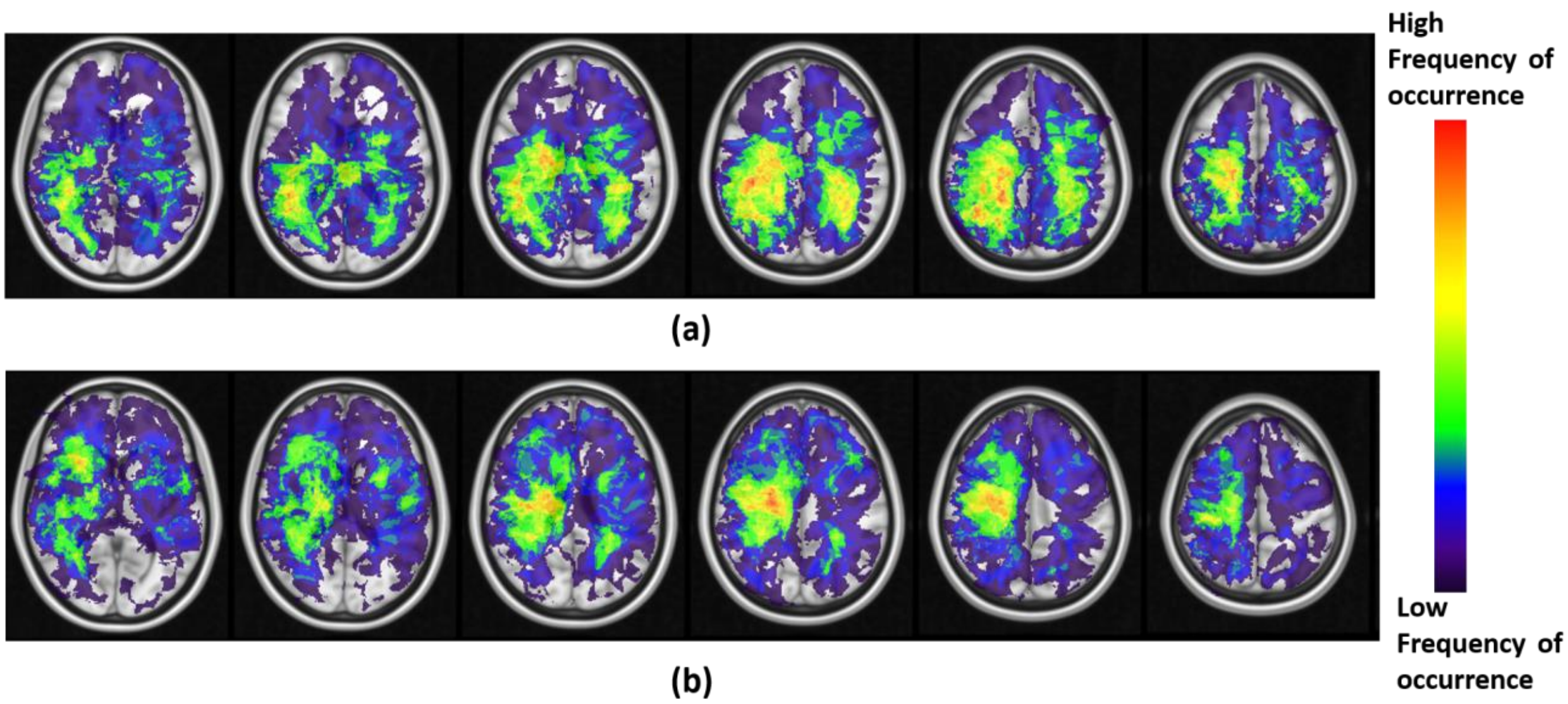

**Figure 2:** (a) Frequency maps of tumor occurrence for peri-lesional T2/FLAIR hyperintensities in tumor recurrence of pre-treatment scans, where lobar prominence is present in the parietal lobe of both hemispheres. (b) Frequency maps of tumor occurrence for peri-lesional T2/FLAIR hyperintensities in pseudo-progression, where a multi-focal distribution is present in the frontal lobe, temporal lobe, insula, and putamen of the right hemisphere.

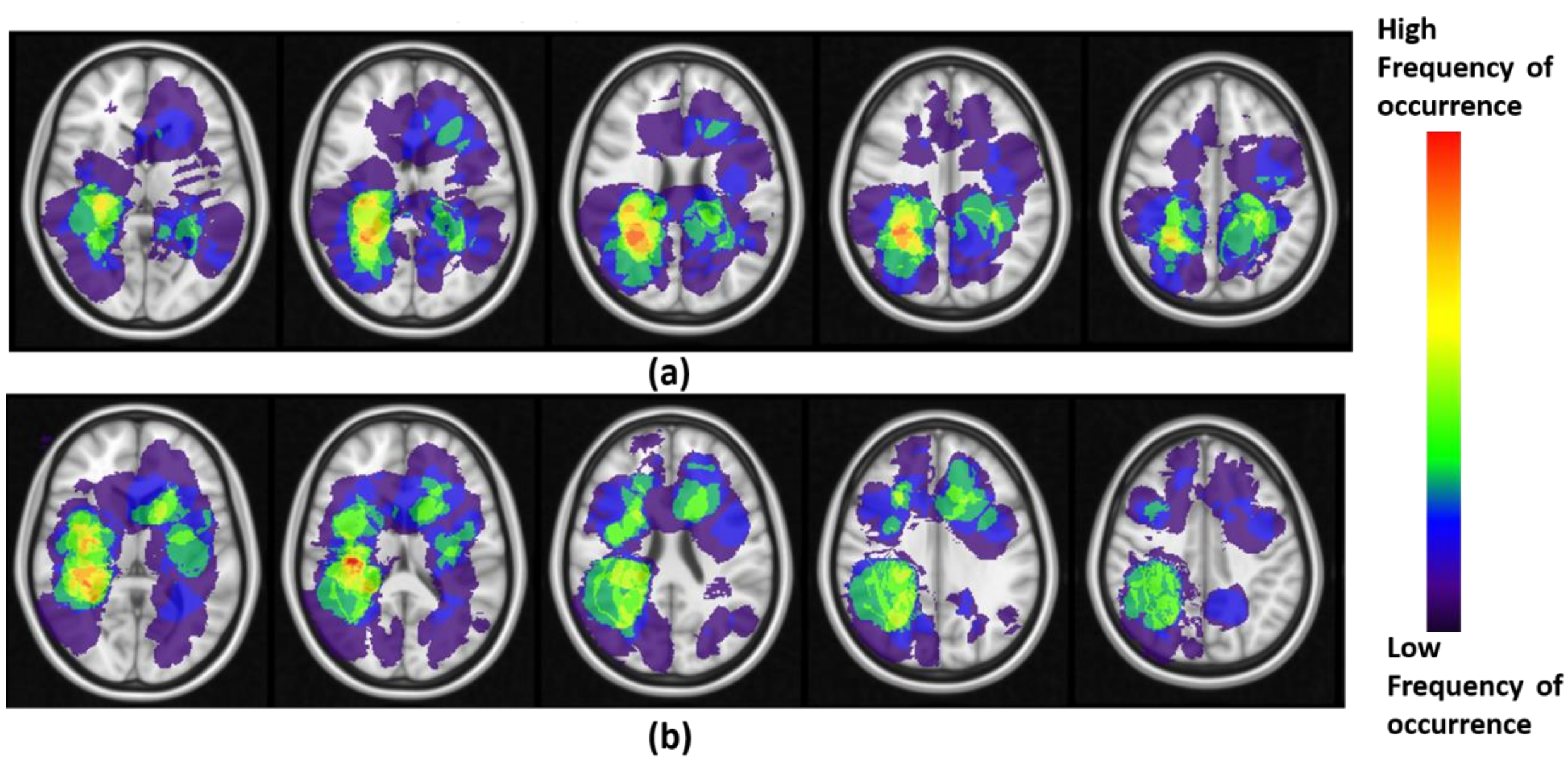

**Figure 3:** (a) Frequency maps of tumor occurrence for enhancing lesion in tumor recurrence of pre-treatment scans, where lobar prominence is present in the parietal lobe of both hemispheres. (b) Frequency maps of tumor occurrence for enhancing lesion in pseudo-progression, where a multi-focal distribution is present in the insula, frontal lobe, putamen, and the temporal lobe, of both left and right hemispheres.

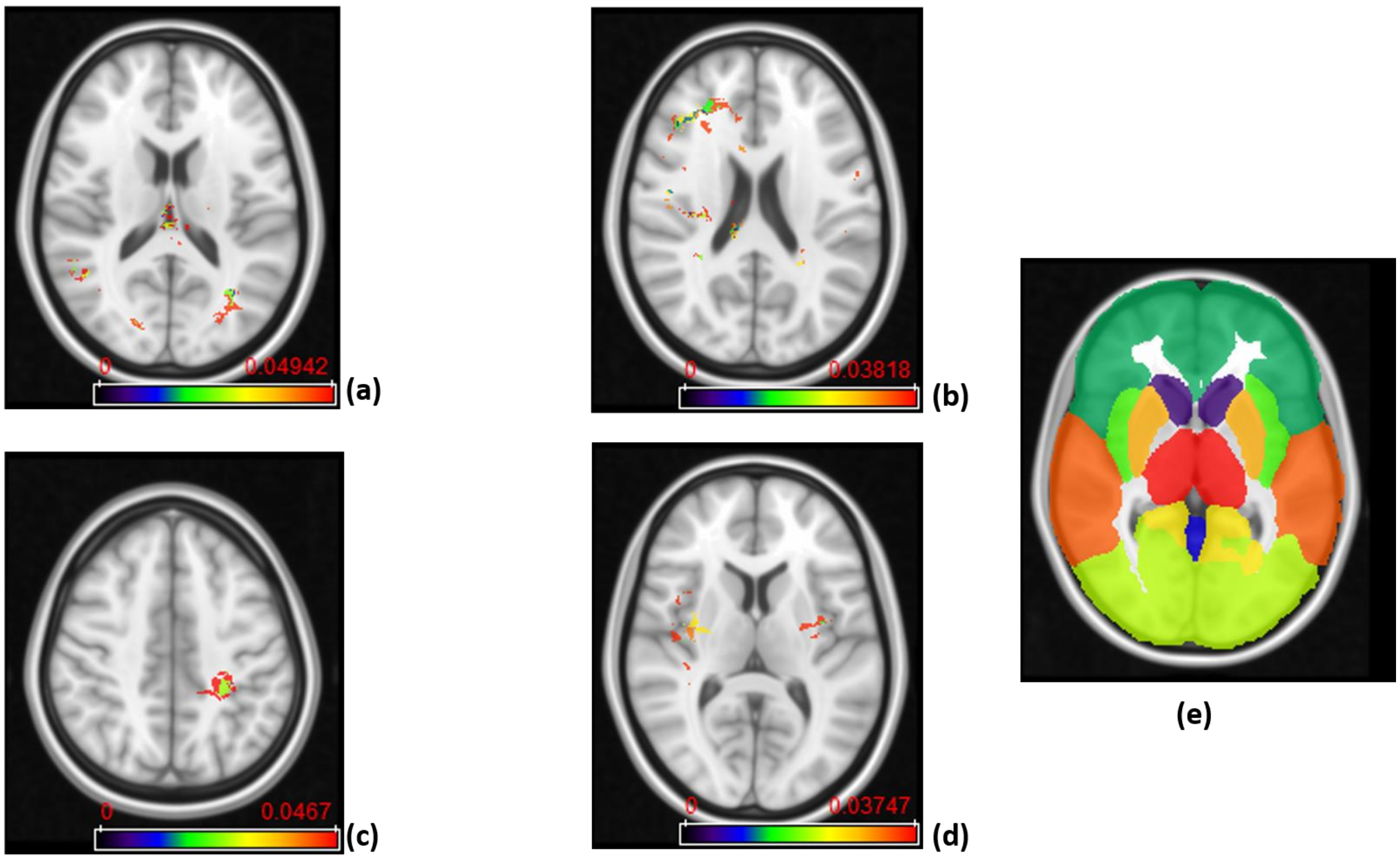

**Figure 4:** (a) ADIFFI maps for peri-lesional T2/FLAIR hyperintensities in tumor recurrence, and (b) pseudo-progression. (c) ADIFFI maps for enhancing lesion in tumor recurrence, and (d) pseudo-progression. The level of significance was at a P-value of 0.05 for all of these maps. These were the maps prior to applying RP analysis. (e) The labeled anatomical MNI atlas that is used for parcellating ADIFFI maps and identifying significant areas.

**Tables**

| Characteristic | Tumor Recurrence | Pseudo-progression |
|---|---|---|
| No. of patients | 41 | 33 |
| Females | 16 | 12 |
| Males | 25 | 21 |
| Mean age (year) | 59.1 | 61.96 |
| Age range (year) | 26 - 75 | 24 - 75 |

**Table 1**: Summary of the study population used in this work to create population atlases for PsP and tumor recurrence.